\documentclass[pra,twocolumn,nofootinbib,preprintnumbers,superscriptaddress,aps]{revtex4-2}

\usepackage{natbib}
\usepackage{amssymb,amsbsy,amsmath,amsfonts}
\usepackage{mathrsfs}
\usepackage{graphicx}% Include figure files
\usepackage{epsf,epsfig,float,latexsym,amsthm,fancyhdr,rotating}
\usepackage{graphics,psfrag,longtable}
\usepackage{slashed}
\usepackage{esint}
\usepackage{nicefrac}
\usepackage{braket}
\usepackage{mathtools}
\usepackage{ulem}

\newcommand{\nn}{\nonumber}

\newcommand{\be}{\begin{equation}}
\newcommand{\ee}{\end{equation}}

\usepackage{color}
\definecolor{purple}{rgb}{.36,.12,.60}
\definecolor{orange}{rgb}{.9,.3,.0}
\definecolor{green}{rgb}{0,.6,0}
\definecolor{RED}{rgb}{.9,.3,.1}

\usepackage[colorlinks,citecolor=blue,linktoc=all,linkcolor=cyan]{hyperref}

\begin{document}

\title {Spin Asymmetries for Twisted-Neutron Capture by Polarized \texorpdfstring{$^3$He}{3He} Nuclei}

\author{Rakhee Roy}

\affiliation{Department of Physics,
The George Washington University, Washington, DC 20052, USA}

\author{Carl E. Carlson}

\affiliation{Physics Department, William \& Mary, Williamsburg, Virginia 23187, USA}

\author{Andrei Afanasev}

\affiliation{Department of Physics,
The George Washington University, Washington, DC 20052, USA}

\begin{abstract}
We provide a theoretical analysis of a nuclear reaction caused by twisted neutrons that carry an additional Orbital Angular Momentum (OAM) along the propagation direction. Angular momentum selection rules are derived for the thermal-neutron capture reaction on a spin-polarized $^3$He nucleus. We demonstrate that noticeable novel spin effects due to OAM are limited to the region of the size of de Broglie wavelength around the phase singularity of the beam. Spin asymmetries are predicted for a range of experimental constraints.
\end{abstract}
\date{\today
}
\maketitle

\pagestyle{plain}
%%%%%%%%%%%%%%%%%%%%%%%%%%%%%%%%%%

\section{Introduction}

%%%%%%%%%%%%%%%%%%%%%%%%%%%%%%%%%%

%Doubters~\cite{2018PhRvL.120i0402C} of neutron twisted beams have showed their existence could be verified by finding certain cross section asymmetries~\cite{Jach:2021veo}, which now have been sought and not found~\cite{sarenac2026experimental}.  Blame has fallen not on the absence of twistedness but on unanticipated severe beam and target localization requirements~\cite{sarenac2026experimental}.  

%The idea that twistedness can reveal itself by exciting transitions with quantum number changes impossible for plane waves is correct.  The idea works in its pure form only if the target is precisely on the vortex axis of the twisted neutron.  When the target is off axis, the amplitudes for the unique quantum number changes possible for twisted incoming projectiles mix with amplitudes that are identical to what can be produced by plane wave beams.  In fact, averaging the results for target placements that are experimentally necessarily spread around the vortex line finds that the plane wave allowed transitions are in overall probability significantly larger than transitions that can only occur for twisted incoming neutrons.

%%%%%%%%%%%%%%%%%%%%%%%%%%%%%

Twisted quantum waves of the thermal neutrons that carry an additional Orbital Angular Momentum (OAM) in the propagation direction were successfully generated and studied experimentally \cite{clark2015,sarenac2019,sarenac2022experimental}.
Beams carrying OAM -- or quantum vortex beams -- have a potential to become a new tool for nuclear physics studies, see, $e.g.$ \cite{ivanov2026vortex} for a recent review.

Several theoretical studies demonstrated novel effects associated with the twisted neutrons in nuclear reactions: Radiative capture of neutrons by a proton \cite{afanasev2018radiative}, longitudinal spin asymmetries of neutron's Schwinger scattering on nuclei that are parity-forbidden for plane waves \cite{afanasev2019schwinger,afanasev2021elastic}, altered angular distributions of the twisted-neutron beta-decay products \cite{kou2025unveiling} and, most recently, use of OAM for partial-wave analysis of nucleon-nucleon scattering \cite{hu2026vortex}.

An interesting opportunity for the capture of twisted neutrons by $^3$He nuclei was pointed out in Ref.\cite{jach2022method} that predicted, for example, reversal of the sign for the longitudinal beam-target asymmetry due to neutron OAM. However, the predicted effect \cite{jach2022method} was experimentally ruled out \cite{sarenac2026experimental} at a per cent level. This theory-experiment controversy motivated the present study.

We derive angular-momentum selection rules for absorption of twisted neutrons by an atomic nucleus, but our result equally applies to twisted-fermion absorption by quantum systems that conserve angular momentum, $e.g.$, capture of twisted electrons by atoms. Similar selection rules were previously derived theoretically \cite{afanasev2013off, scholz2014absorption} and demonstrated experimentally \cite{schmiegelow2016transfer,afanasev2018experimental} for the absorption of twisted photons by atoms. %Here, we derive matrix elements of quantum transitions induced by twisted spin-1/2 particles and demonstrate the effect of OAM on spin asymmetries for the absorption of twisted neutrons by polarized $^3$He. 
Using the developed formalism for a specific nuclear reaction, namely, the capture of thermal twisted neutrons by polarized $^3$He nuclei, we clarify the origins of the null effect  seen in experiment \cite{sarenac2026experimental}. 

%%%%%%%%%%%%%%%%%%%%%%%%%%%%

\section{Capture Amplitudes for Twisted Neutrons}

%%%%%%%%%%%%%%%%%%%%%%%%%%%%%
%\subsection{Plane-wave capture amplitude}

We consider twisted neutron projectiles and, when we are specific, the reaction 
$n\, + ^3$He $\to ^4$He$^*$.  We will use Bessel states for the twisted neutrons, and consider both the case that the neutron waves forming the twisted state all have the same helicity and, separately, the case that the neutrons are all polarized along the $z$-direction, the direction of propagation. We opt for the Bessel model as opposed to Laguerre-Gauss (LG) for the twisted-neutron states based on the experience with twisted photons \cite{afanasev2018experimental}: Selection rules for them are essentially the same for the entire beam profile through the first intensity peak ($i.e.$, the ``doughnut" radius), while Bessel states provide simpler analytic solutions for the transition matrix elements.

For Bessel states with definite neutron helicity,
\be
\label{eq:helstate}
\ket{\psi_{\kappa m_n k_z \Lambda} } = e^{i k_z z}\int \frac{ d\phi_k}{ 2\pi } (-i)^{m_n} e^{im_n \phi_k} 
	e^{i \vec k \cdot \vec b}	\ket{ \vec k, \Lambda }
\ee
where $m_n$ is total angular momentum (TAM) projected along $z$-direction. States $\ket{\vec k, \Lambda}$ are momentum eigenstates of helicity $\Lambda$.  For the Bessel state, all $\vec k$ lie on the surface of a cone in momentum space, having all the same polar angle $\theta_k$ (the pitch angle) and varying azimuthal angles $\phi_k$. One has $k_z 
= | \vec k| \cos\theta_k = k \cos\theta_k$ and $\kappa = k \sin\theta_k$.  Impact parameter $\vec b$ is the transverse displacement of the vortex line of the twisted beam from the $z$-axis, where the target state will be centered.   For the record, expectation values of the projected spin angular momentum (SAM) and orbital angular momentum (OAM) are
$\braket{S_z} =\Lambda \cos\theta_k$ and $\braket{L_z} = m_n - \Lambda\cos\theta_k$~\cite{afanasev2013off}.

In the position representation, choosing quantization along $z$-axis, integration (\ref{eq:helstate}) yields the wave function of twisted neutron state,
\begin{align}
\psi_{\kappa m_n k_z \Lambda}=e^{i k_z z}&\sum_\sigma i^{-\sigma} J_{m_n-\sigma}(\kappa b) \\ \nonumber 
\times &e^{i(m_n-\sigma) \phi_b} d^{1/2}_{\sigma\Lambda}(\theta_k)\chi_\sigma,
\end{align}
where $d^{1/2}_{\sigma\Lambda}$ are Wigner rotation matrices and the two-component spinors are:
\be
\chi_{\frac{1}{2}}=\begin{pmatrix} 1\\ 0   \end{pmatrix}, \ \ \chi_{-\frac{1}{2}}=\begin{pmatrix} 0\\ 1   \end{pmatrix}. 
\ee
States having the same momenta but having a definite spin projection $s$ along $z$-axis are
\be
\label{eq:spinVShel}
\ket{ \vec k, s} = e^{i s \phi_k } \sum_\Lambda d^{1/2} _{s,\Lambda} (\theta_k) \ket{ \vec k, \Lambda },
\ee
and twisted states of definite spin along $z$-axis are
\be
\ket{ \hat \psi_{\kappa \ell k_z s} } = e^{i k_z z}\int \frac{ d\phi_k}{ 2\pi } (-i)^{\ell+s} e^{i \ell \phi_k} 
	e^{i \vec k \cdot \vec b}	\ket{ \vec k, s }.
\ee

The corresponding position-space wave function is
\begin{align}
\psi_{\kappa l k_z s}=e^{i k_z z}& i^{-s} J_{l}(\kappa b)
 e^{i l \phi_b} \chi_\sigma.
\end{align}

For the expectation values of spin and OAM operators, we find: 
\be
\braket{S_z} = s, \quad \braket{L_z} = \ell,  \quad \text{and} \quad \braket{J_z} = \ell + s .
\ee

Transition amplitudes for twisted neutrons of definite $\Lambda$ are given by
\begin{align}
M^{ (tw) }_{m_f m_i m_n \Lambda}( \vec b ) = \braket{ \beta j_f m_f | H_{int} | 
	\alpha j_i m_i ; \psi_{\kappa m_n k_z \Lambda}	}
\end{align}
where $\{j_i,m_i\}$ are (for a general case) the spin and its projection along the $z$-axis for the initial state and $\alpha$ represents other information needed to identify the initial nuclear state, with 
$\{j_f,m_f\}$ and $\beta$ being the corresponding quantities for the final state.  One may proceed similar to Refs.~\cite{scholz2014absorption,afanasev2018experimental,2019JOSAB..36..565S} for twisted photons and relate this amplitude to a transition amplitude for plane-wave neutrons and factors that depend only on beam's properties and target's location:
\begin{align}
\label{eq:helsrule}
&M^{ (tw) }_{m_f m_i m_n \Lambda}( \vec b ) = i^{m_i-m_f} 
e^{i (m_n + m_i - m_f) \phi_b }     \nn\\
&\  \times	
J_{m_f - m_i - m_n}( \kappa b ) 
\sum_{m'_f m'_i} 
    d^{j_f}_{m_f m'_f}(\theta_k)
		d^{j_i}_{m_i m'_  i }(\theta_k)	\,	M^{ (pl) }_{m'_f m'_i  \Lambda} .
\end{align}
The last factor is the plane wave amplitude
\begin{align}
M^{ (pl) }_{m'_f m'_i  \Lambda}( \vec b ) = \braket{ \beta j_f m'_f | H_{int} | 
	\alpha j_i m'_i ; k \hat z, \Lambda	} ,
\end{align}
where all momenta and all spin projections or helicities are along the $z$-axis. One necessarily has
$ m'_f = m'_i + \Lambda.$  Some further indices that could be included in $M^{(tw)}$ and $M^{(pl)}$ have been left tacit.

Before returning to the spin along the $z$-axis case, we define beam polarization asymmetries obtained by comparing transition rates when flipping the helicity of the incident neutron while keeping $m_f, m_i,$ and 
$\ell \equiv m_n - \Lambda$ fixed,
\begin{align}
\label{eq:helBas}
A_{hel}^{(beam)} 
%&= A_{hel}(m_f,m_i,\ell) \nn\\
&=	\frac{ | M^{ (tw) }_{m_f , m_i  , \ell-1/2, -1/2}  |^2 - | M^{ (tw) }_{m_f , m_i  , \ell+1/2, 1/2}  |^2 }
		{ | M^{ (tw) }_{m_f , m_i  , \ell-1/2, -1/2}  |^2 + | M^{ (tw) }_{m_f , m_i  , \ell+1/2, 1/2}  |^2 }  .
\end{align}
The $m_f$ may be fixed, as it is here, or alternatively it may be summed over in the individual terms.

Let us now consider the transition amplitude for the twisted neutron state of definite spin along the $z$-direction is
\begin{align}
\hat M^{ (tw) }_{m_f m_i \ell s}( \vec b ) = \braket{ \beta j_f m_f | H_{int} | 
	\alpha j_i m_i ; \hat\psi_{\kappa \ell k_z s}	}
\end{align}
which leads to
\begin{align}
\label{eq:spinSrule}
&\hat M^{ (tw) }_{m_f m_i \ell s}( \vec b ) = i^{ m_i -m_f} e^{i (\ell + s + m_i-m_f ) \phi_b }
	J_{ m_f - m_i - s - \ell }( \kappa b ) 
                    \nn\\
&\times    \sum_{\Lambda m'_f m'_i} 
    d^{1/2}_{s \Lambda}(\theta_k)
		d^{j_f}_{m_f m'_f}(\theta_k)
			d^{j_i}_{m_i m'_  i }(\theta_k)	\,	M^{ (pl) }_{m'_f m'_i  \Lambda}
\end{align}

The beam asymmetry, for $m_f, m_i, \text{ and } \ell$ fixed and $s$ flipping is
\begin{align}
\label{eq:spinBas}
A_{spin}^{(beam)} 
%&= A_{spin}(m_f,m_i,\ell) \nn\\
&= \frac{ | \hat M^{ (tw) }_{m_f , m_i  , \ell, -1/2}  |^2 - | \hat M^{ (tw) }_{m_f , m_i  , \ell, 1/2}  |^2 }
		{ | \hat M^{ (tw) }_{m_f , m_i  , \ell, -1/2}  |^2 + | \hat M^{ (tw) }_{m_f , m_i  , \ell, 1/2}  |^2 }
\end{align}
Again, one can choose between having $m_f$ fixed or summing $m_f$ in the individual terms.

As a simple example take $j_f = m_f = 0$.  For the helicity case,
\begin{align}
M^{ (tw) }_{0 m_i m_n \Lambda}( \vec b ) &= i^{m_i} e^{i (m_n + m_i ) \phi_k }
J_{- m_i - m_n}( \kappa b ) 	
                \nn\\
&\quad \times			d^{j_i}_{m_i, - \Lambda }(\theta_k)	\,	M^{ (pl) }_{0,-\Lambda, \Lambda}	\,.
\end{align}
For neutron spin along the $z$-axis and the present case of $j_i=1/2$, and further with parity giving 
$M^{ (pl) }_{0, -\Lambda,  \Lambda} 
= M^{ (pl) }_{0, \Lambda,  -\Lambda} $,  obtain
\begin{align}
\label{eq:spinSrule}
\hat M^{ (tw) }_{0 m_i \ell s}( \vec b ) &= i^{m_i} e^{i (\ell + s + m_i ) \phi_b }
	J_{- m_i - s - \ell}( \kappa b )
                \nn\\   &\quad\times
		d^{1/2}_{m_i, - s }(2 \theta_k)	\,	M^{ (pl) }_{0,\mp1/2, \pm 1/2}	\,.
\end{align}
(Note that summation over $\Lambda$ leads to doubling the argument of the Wigner function.)

It is important to emphasize that Eqs.~(\ref{eq:helsrule},\ref{eq:spinSrule}) include a plane-wave transition amplitude as a factor. It follows that if a state with a given total angular momentum $j_f$ is forbidden for plane waves, it would remain forbidden for twisted incident waves.  

Staying with $j_f = 0$, and specifying 
$\ell\ge 1$ and  
$m_i = -1/2$, we get
\begin{align}
A_{hel}^{(beam)} 
%&= A_{hel}(0,-1/2,l) \nn\\
&=	\frac{ J_{l-1}^2(\kappa b) \sin^2(\theta_k/2) -  J_l^2(\kappa b) \cos^2(\theta_k/2) }
		{ J_{l-1}^2(\kappa b) \sin^2(\theta_k/2) +  J_l^2(\kappa b) \cos^2(\theta_k/2) }
\end{align}
and
\begin{align}
A_{spin}^{(beam)} 
%&= A_{spin}(0,-1/2,l) \nn\\
&= 
	\frac{  J_{l-1}^2(\kappa b) \sin^2\theta_k  -  J_l^2(\kappa b) \cos^2\theta_k }
		{ J_{l-1}^2(\kappa b) \sin^2\theta_k  +  J_l^2(\kappa b) \cos^2\theta_k }		\,.
\end{align}

Other spin asymmetries are target asymmetries that may be formed by fixing the values of $(l,\Lambda)$ or $(l,s)$, while comparing the reaction rates for opposite signs of $m_i$:

\begin{align}
A_{hel}^{(tar)} &=	\frac{ | M^{ (tw) }_{m_f ,-1/2 , \ell+\Lambda, \Lambda}  |^2 - | M^{ (tw) }_{m_f , 1/2  , \ell+\Lambda, \Lambda}  |^2 }
		{ | M^{ (tw) }_{m_f ,-1/2 , \ell+\Lambda, \Lambda}  |^2 + | M^{ (tw) }_{m_f , 1/2  , \ell+\Lambda, \Lambda}  |^2 } \, \\
A_{spin}^{(tar)} 
&= \frac{ | \hat M^{ (tw) }_{m_f , -1/2  , \ell, s}  |^2 - | \hat M^{ (tw) }_{m_f , 1/2 , \ell, s}  |^2 }
		{ | \hat M^{ (tw) }_{m_f , -1/2  , \ell, s}  |^2 + | \hat M^{ (tw) }_{m_f , 1/2 , \ell, s}  |^2 }        
\end{align}

Taking again $j_f=0$ for example,  we arrive at the expressions for target asymmetries:
\begin{align}
A_{hel}^{(tar)} &=\frac{ J_{l+2\Lambda}^2(\kappa b) \sin^2(\theta_k/2) -  J_l^2(\kappa b) \cos^2(\theta_k/2) }
		{ J_{l+2\Lambda}^2(\kappa b) \sin^2(\theta_k/2) +  J_l^2(\kappa b) \cos^2(\theta_k/2) } \\
A_{spin}^{(tar)} 
&=\frac{ J_{l+2\Lambda}^2(\kappa b) \sin^2(\theta_k) -  J_l^2(\kappa b) \cos^2(\theta_k) }
		{ J_{l+2\Lambda}^2(\kappa b) \sin^2(\theta_k) +  J_l^2(\kappa b) \cos^2(\theta_k) }.     
\end{align}
Assuming $l>0$, we see that the target asymmetries coincide with the corresponding beam asymmetries for $\Lambda=-1/2$ or $s=-1/2$. If, on the other hand, $\Lambda=1/2$ or $s=1/2$, then the target and beam asymmetries are different.

In the paraxial limit $\theta_k\ll1$, we can use Taylor expansion for the Bessel functions, $J_n(x)\approx (\frac{x}{2})^l\frac{1}{l!}$, resulting in simplified expressions: 
\begin{align}       \label{eq:exp}
&A_{hel}^{(tar)}|_{l\cdot\Lambda<0}=A_{hel}^{(beam)}=\frac{l^2-(kb)^2}{l^2+(kb)^2},\\
&A_{spin}^{(tar)}|_{l\cdot\Lambda<0}=A_{spin}^{(beam)}=\frac{4l^2-(kb)^2}{4l^2+(kb)^2}, \\
&A_{hel}^{(tar)}|_{l\cdot\Lambda>0}=A_{spin}^{(tar)}|_{l\cdot\Lambda>0}=-1.
\end{align}

%%%%%%%%%%%%%%%%%%%%%%%%%%%%%%

\section{Numerical Analysis}

%%%%%%%%%%%%%%%%%%%%%%%%%%%%%%

Consider the reaction $n\, + ^3$He $\to ^4$He$^*$ that was recently studied with a twisted-neutron beam \cite{sarenac2026experimental}. For thermal plane-wave neutrons, the reaction is known to proceed via formation of a compound $^4$He$^*$ nucleus with zero spin and positive parity \cite{passell1966measurement}. Therefore, for the plane waves, absorption takes place for opposite spins of the incident neutron and $^3$He target and the asymmetries defined above would always be $-1$.  For the twisted wave, the asymmetry is $+1$ when the target is centered on the vortex axis, $i.e.$, for $b=0$,  but quickly goes to near $-1$ and remains there for nearly all values of $b$. 

\begin{figure}[t!]
\includegraphics[width=0.95\columnwidth]{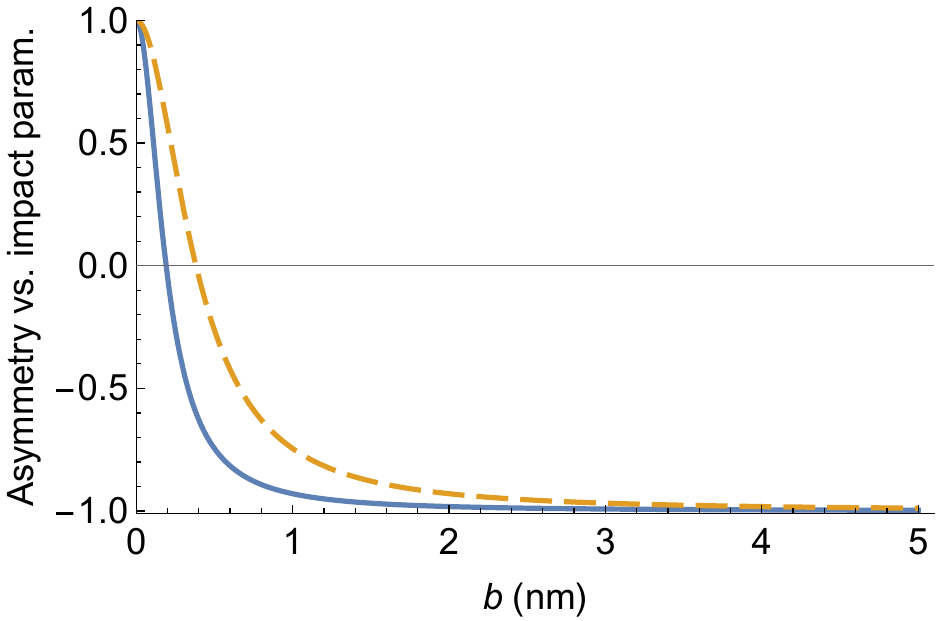}
\caption{Asymmetries for $\ell = 1$ with flipping the beam helicity Eq.(\ref{eq:helBas}) (solid blue curve) or flipping the spin along the $z$-axis Eq.(\ref{eq:spinBas}) (dashed gold curve).  This figure is for $\lambda = 1.2$ nm and $\theta_k = 3.8 \times 10^{-7}$ radians.  The abscissa of this figure ends before reaching a point where $J_1$ is zero other than at $b=0$.}
\label{fig:asymmvsb}
\end{figure}

We chose neutron's wavelength $\lambda=1.2$~nm and $\theta_k$=3.8$\times$10$^{-7}$ radians to match the beam energy and a 1.0~mm beam radius of Ref.\cite{sarenac2019} with the Bessel model used here. (Namely, this value of $\theta_k$ places the maximum of beam's intensity at $b=$1.0~mm from the central axis). Fig.~\ref{fig:asymmvsb} shows the asymmetry with $\ell = 1$ for the helicity case (solid blue curve) and spin along the $z$-axis case (dashed gold curve). The exception is a target asymmetry for same-sign spin $s$ or helicity $\Lambda$ and OAM $l$ of incident neutrons: In this case, the asymmetry is the same as for plane waves.   (Given the expansion of the Bessel functions for small arguments and the definition of $\kappa$, both terms in the numerator have the same leading $\theta_k$ dependence, so the falloff of the asymmetry with $b$ is only weakly pitch angle dependent, as in Eqs.~\eqref{eq:exp}ff.)

%%%%%

%%%%%%%

Generally one cannot expect the target location to be well localized relative to the vortex line of the beam.  To see the effect of a spread out target, consider a target localized within a disc of radius $b_\text{disc}$, with uniform target density with this radius and zero density outside.  The asymmetries should then be redefined with the individual terms integrated over the area of the disc before taking the ratios.  A useful integral is
\be
\mathcal I_n(x) = \int_0^x J^2_n(x') 
    \, x' dx'		\,,
\ee
which can be integrated analytically and also expanded in $x$ for large$x$.  Again taking $\ell \geq 1$ and $j_f = 0$ as an example, the integrated helicity based asymmetry becomes
\begin{align}
\label{eq:intHelBas}
&A_{hel}^{(beam,int)}           \nn\\
&\ =      \frac{	
\mathcal I_{\ell-1}(\kappa b_\text{disc})
    \sin^2(\theta_k/2)
- \mathcal I_\ell(\kappa b_\text{disc}) \cos^2 (\theta_k/2)}  {
\mathcal I_{\ell-1}(\kappa b_\text{disc})
    \sin^2(\theta_k/2)
+ \mathcal I_\ell(\kappa b_\text{disc}) \cos^2 (\theta_k/2)}		\,,
\end{align}
with the spin case the same but for 
$\theta_k/2 \to \theta_k$.  The integrated asymmetry remains $+1$ for very small $b_\text{spot}$. For large $b_\text{disc}$,  we can use the large argument expansion, which has a leading term that grows monotonically plus oscillating corrections,
\begin{align}       \label{eq:expand}
\mathcal I_n(x) &= \frac{x}{\pi} - \frac{1}{2 \pi}
	\sin \left( 2x-(1+2n)\frac{\pi}{2} \right)         \nn\\
&\,+ \frac{1}{x \pi }  \sin^2 \left( x-(1+2n)\frac{\pi}{4} \right)
			\left( \frac{1}{4} - n^2 \right) + \ldots
\end{align}
For the present situation, one needs only the leading term.
Hence for targets spread over a large disc,
\begin{align}
A_{hel}^{(beam,int)} &= -\cos^2 \theta_k + 
    \mathcal{O}(1/(\kappa b_\text{spot})),
    \quad \text{or}     \nn\\
A_{spin}^{(beam,int)} &= -\cos^2( 2\theta_k) + \mathcal{O}(1/(\kappa b_\text{spot})),
\end{align} 
i.e., essentially $-1$ for small pitch angle.

\begin{figure}[h]
\includegraphics[width=0.95\columnwidth]{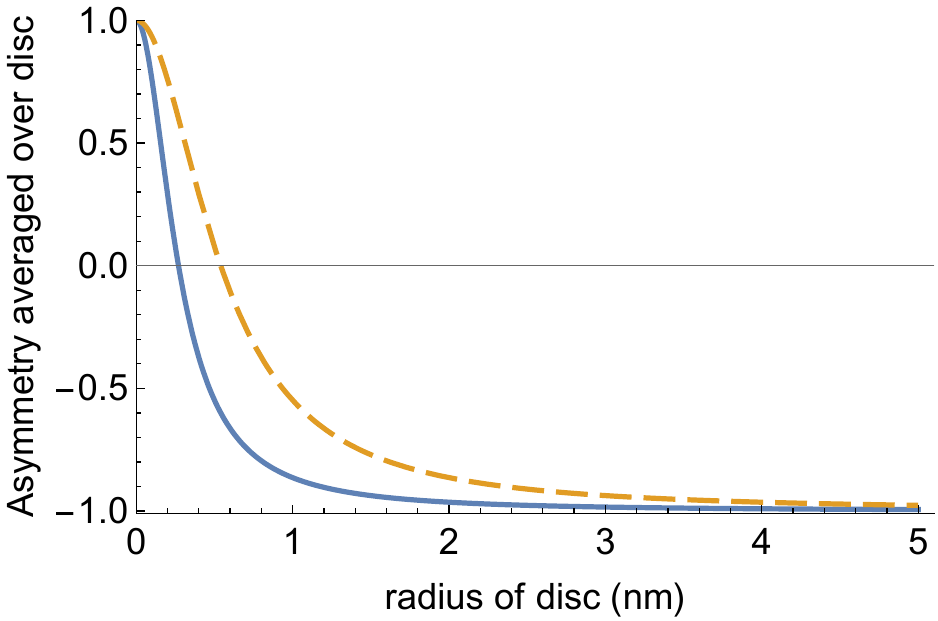}
\caption{Asymmetries integrated over a target disc, Eq.(\ref{eq:intHelBas}) for $\ell = 1$, with flipping the beam helicity (solid blue curve) or flipping the spin along the $z$-axis (dashed gold curve).  Beam parameters are as in Fig. 1. }
\label{fig:int}
\end{figure}

A plot of the integrated asymmetries looks rather similar to the $b$-specific asymmetries, Fig.~\ref{fig:int}, except that the denominator never gets small at any value of $b_\text{disc}$ not very small.

%%%%%

%%%%%%%

{%\color{red}

%%%%%%%%%%%%%%%%%%%%%%%%%%%%%%%%%%

\section{Theory-Experiment Comparison}

%%%%%%%%%%%%%%%%%%%%%%%%%%%%%%%%%%%

%{\large NOTE TO US: For a short paper we may not want or need to review the Sarenac et al asymmetry choice.  It could be included or it could go in some longer paper later.  \bigskip}

Sarenac et al~\cite{sarenac2026experimental} define and measure cross section differences, in their notation,
\be
\delta\sigma_{\ell} = \sigma_{\ell\uparrow\downarrow} - \sigma_{\ell\uparrow\uparrow}
\ee
where the indices represent $\sigma_{\ell m_i s}$, and we will connect to our notation momentarily.  We treat the case that the incoming neutrons have spin aligned or anti-aligned along the $z$-axis.  (We could also do the definite helicity case, with the indices representing 
$\sigma_{\ell m_i \Lambda} = \sigma_{m_n-\Lambda, m_I, \Lambda}$.)  The dependence on impact parameter $b$ is here tacit.  The experiments did not measure quantities in the final state, so that $j_f$ and $m_f$ are de facto summed over.  Restricting to a specific $j_f$ and $m_f$,  the connection to our notation is
\be
\sigma_{\ell m_I s} \propto | \hat M^{(tw)}_{m_f m_i \ell s} |^2		\,,
\ee
with $m_f = m_i + \ell + s$ if on axis.

The asymmetries defined in Sarenac et al are
\be
\Delta\sigma(\ell_1,\ell_2) = 
\frac{ \delta\sigma_{\ell_1} - \delta\sigma_{\ell_2} }{ \delta\sigma_{\ell_1} + \delta\sigma_{\ell_2}  }	\,.
\ee
Note that although the character $\sigma$ is used, this is an asymmetry.  Also note that $\delta\sigma_\ell$ can go negative, so that this particular asymmetry could in some circumstances have magnitude larger than one.

These asymmetries are designed to see if the twistedness, measured by $\ell_{1}$ or $\ell_{2}$, has an effect on the cross sections.  If the twistedness, coming from the orbital angular momentum, is irrelevant, these asymmetries would be zero.  

As an example, we work out the asymmetry for $\ell_1=-1$ and $\ell_2= 1$, with, as noted above, $m_i = 1/2$ and further constraining $j_f = m_f = 0$.  One obtains
\begin{align}
&\Delta\sigma_b(-1,1)       \nn\\
&\ = \frac{ \left( 	J_2^2(\kappa b) - J_0^2(\kappa b)	\right) \sin^2 \theta_k }
{ 2 J_1^2(\kappa b) \cos^2 \theta_k - \left( 	J_2^2(\kappa b) + J_0^2(\kappa b)	\right) \sin^2 \theta_k }
\end{align}
This is the asymmetry at a given impact parameter $b$.  The asymmetry is 100\% in magnitude at zero impact parameter, and significantly smaller for most non-zero $b$.

(For interest, we can note that if we had used the definite helicity states with $\Lambda = \pm 1/2$ instead of the $s = \pm 1/2$ states, but still using $\ell = \pm 1$, we would obtain the same result but with the pitch angle $\theta_k$ replaced by $\theta_k/2$.)

Generally,  the target as seen by the beam is spread over some area.   In this case, one can integrate the individual terms in the expression above over the target area.  We will again show results, for definiteness, for the case that the target density is uniform within a disc of radius $b_\text{disc}$ and zero beyond that.  The relevant integrals can be done analytically and the definitions and large argument expansions have been given earlier, in Eq.~\eqref{eq:expand}.  We note that at large $b$ (say on a scale judged by the wavelength of the incoming neutron) there is cancellation in the numerator and buildup in the denominator, abetted by the $\theta_k$ factors, leading to remarkably small expected asymmetries.

Using the earlier notation, and for the disc conditions just noted, the scattering asymmetry for the same quantum numbers as above is
\begin{align}
\label{eq:deviant}
&\Delta\sigma(-1,1)     \nn\\
&= \frac{ \left( 	
\mathcal I_2(\kappa b_\text{disc}) - \mathcal I_0(\kappa b_\text{disc})	\right) \sin^2 \theta_k }
{ 2 \mathcal I_1^2(\kappa b_\text{disc}) \cos^2 \theta_k -
\left( \mathcal I_2^2(\kappa b_\text{disc}) + \mathcal I_0^2(\kappa b_\text{disc})	\right) \sin^2 \theta_k }		\,.
\end{align}

Using the expansions, the very large $b$ result is
\be
\Delta\sigma(-1,1) \approx \frac{ 2  \sin^2 \left( \kappa b_\text{disc} - \frac{\pi}{4} \right) }
	{ (k b_\text{disc})^2 \cos(2\theta_k)  }		\,,
\ee
where we have recalled the definition of $\kappa$, so that the dependence of the envelope of the falloff on pitch angle is slight, even while the period of the oscillations with the disc radius depends significantly on the pitch angle

\begin{figure}[h]
\includegraphics[width=0.95\columnwidth]{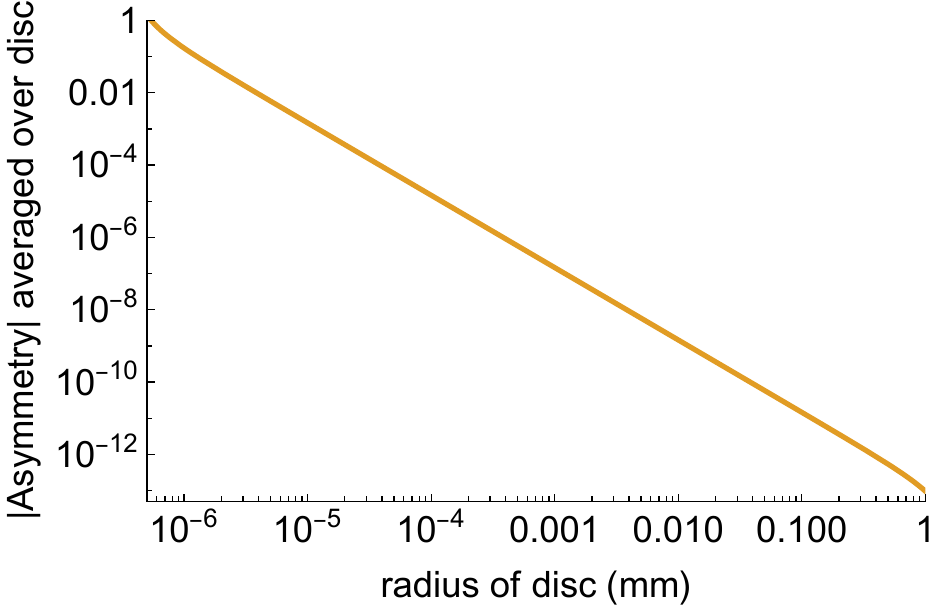}
\caption{       
OAM-induced deviation of spin asymmetry integrated over a target disc, $\Delta\sigma(-1,1)$, as defined in Eq.(\ref{eq:deviant}).  Beam parameters are as in Fig.1.
   }      
\label{fig:asympsarenac}
\end{figure}

Analytic results for $\Delta\sigma(-1,1)$ are seen  in Fig.~\ref{fig:asympsarenac}.  (Oscillations lie at higher values of the disc radius, for the pitch angle we are using.)   The scale on the horizontal axis is different from our previous plots.  Here we are emphasizing the large target size effect on the falloff of the observable asymmetry, before we were showing how quickly the transition to nearly the plane wave result occurred.  The results shown here are very small at large $b_\text{disc}$ because of cancellations in the numerator and result lies under an envelope that falls quadratically with disc radius.  			

It follows from Fig.~\ref{fig:asympsarenac} that the calculated deviation from the plane-wave asymmetry under experimental conditions of Ref.\cite{sarenac2026experimental} is of the order 10$^{-13}$, far below the experimental accuracy of about 1\%. Further analysis of Eq.(\ref{eq:deviant}) shows that the effect of OAM on the spin asymmetry scales with the square of pitch angle $\theta_k^2$: Choosing $\theta_k$= 0.1 (0.3) rad, we obtain the order of 1\% (10\%) absolute correction. These values of $\theta_k$ imply strong focusing of the beam down to the spot size of a few wavelengths -- to be compared against a mm-size of the recent experiment \cite{sarenac2026experimental} that corresponds to about 10$^5$ neutron's wavelengths.
%-- presenting a significant experimental challenge.
%%%%%

%%%%%%%

%}

%\setcounter{section}{3}

%%%%%%%%%%%%%%%%%%%%%%%%%%%%%%%%%%

\section{Summary and Conclusions}

%%%%%%%%%%%%%%%%%%%%%%%%%%%%%%%%%%%

Twisted neutrons can generate transitions that plane waves cannot.  The manifestations of this are most dramatic when the target is well centered on the twisted beam's vortex line.  However, twisted neutrons are also capable of copiously generating the same transitions that plane waves can.  This will happen when the target is well off the vortex line of the beam.  Seeing or not seeing the unique transitions that twisted beams can deliver becomes a question of how close, in terms of wavelengths, the target center can be to the beam's vortex line, quantified by the selection rules derived here (\ref{eq:helsrule}, \ref{eq:spinSrule}) for different examples of the twisted beam's polarization. 

An experiment was done recently ~\cite{sarenac2026experimental} with twisted thermal neutrons incident on polarized $^3$He nuclei to seek transitions with quantum number changes that are impossible for plane waves~\cite{jach2022method}.  The experiment found no evidence of such transitions.  The authors of Ref.~\cite{sarenac2026experimental} suggested correctly, according to the present calculation, that this was due to the large average separation between the target nuclei and the vortex axis of the beam.

A strong goal of this note has been to quantify how large or how small certain asymmetries, that could be signals of the special quantum number dependencies allowed by twisted beams, could be.  We found that these asymmetries were significant, of the order of unity, at zero impact parameter, $i.e.$, with the target nucleus right on the vortex of the beam, but had relaxed most of the way to their plane wave values as the target-to-vortex line separation had reached one neutron wavelength and higher.  

When considering that the nucleon targets could be in a cell of significant area facing the beam, one needs to average the corresponding cross sections with respect to a range of possible impact parameters. For bulk targets, we find that integrated asymmetries with remarkably small deviations from the plane wave expectation are of order (logarithmically) about $10^{-13}$ for twisted thermal neutrons with a mm beam spot. 

To optimize experiments for observation of larger OAM-induced corrections to spin asymmetries, one would require either restricting the measurement to a target sample of a wavelength size or, alternatively, strong focusing of the twisted beams down to the wavelength size spots.
\medskip

\section*{Acknowledgements}

A.A. thanks DOE for support under the program “Quantum Horizons: QIS Research and Innovation for Nuclear Science” under grant  DE-SC0023695. C.E.C. thanks NSF for support under grant PHY-1812326. The authors thank N.~Geerits, D.~Pushin, D.~Sarenac, W.M.~Snow and S.~Sponar for useful discussions.

\bibliography{vortneut}

\end{document}